# How do neurons operate on sparse distributed representations? A mathematical theory of sparsity, neurons and active dendrites[1]

**Subutai Ahmad[1]\*, Jeff Hawkins[1]**

[1]Numenta, Inc., Redwood City, CA, USA

**\* Correspondence:** Subutai Ahmad, Numenta Inc., 791 Middlefield Road, Redwood City, CA 94063 USA.
sahmad@numenta.com

Abstract

We propose a formal mathematical model for sparse representations and active dendrites in neocortex. Our model is inspired by recent experimental findings on active dendritic processing and NMDA spikes in pyramidal neurons. These experimental and modeling studies suggest that the basic unit of pattern memory in the neocortex is instantiated by small clusters of synapses operated on by localized non-linear dendritic processes. We derive a number of scaling laws that characterize the accuracy of such dendrites in detecting activation patterns in a neuronal population under adverse conditions. We introduce the union property which shows that synapses for multiple patterns can be randomly mixed together within a segment and still lead to highly accurate recognition. We describe simulation results that provide further insight into sparse representations as well as two primary results. First we show that pattern recognition by a neuron with active dendrites can be extremely accurate and robust with high dimensional sparse inputs even when using a tiny number of synapses to recognize large patterns. Second, equations representing recognition accuracy of a dendrite predict optimal NMDA spiking thresholds under a generous set of assumptions. The prediction tightly matches NMDA spiking thresholds measured in the literature. Our model matches many of the known properties of pyramidal neurons. As such the theory provides a mathematical framework for understanding the benefits and limits of sparse representations in cortical networks.

## 1.    Introduction

A wealth of empirical evidence suggests the neocortex represents information using sparse distributed patterns of activity (Barth and Poulet, 2012). Representations are sparse because at any point in time only a small percentage of neurons are active while the rest are inactive. Representations are distributed because although each active neuron contributes information, it is the set of active neurons that determine what is being represented. The diversity of sparse distributed representations (SDRs) in neocortex is remarkable. Sparse representations exist in early auditory, visual and somatosensory areas (Hromádka et al., 2008; Weliky et al., 2003; Vinje and Gallant, 2000; Crochet et al., 2011). These representations correspond directly to sensory features such as visual

---

edges and audio frequency bands. Representations in higher cortical areas are more abstract and categorical in nature. These areas can encode concepts that have no direct correlation to sensory features (Kiani et al., 2007). Primary motor areas encode sparse motor maps that correspond to specific movements (Graziano and Aflalo, 2007). Premotor areas encode more abstract behavioral "plans" such as the notion of "grasping and inserting food into your mouth" (Graziano et al., 2002). The ubiquity of sparse distributed representations suggests that they enable operations essential for brain function and neural computation. To function effectively these representations must have tremendous capacity and must be extremely tolerant to noise. The exact laws governing their behavior are unknown.

Our experimental understanding of the mechanics of how neurons operate on their inputs has evolved significantly over the last 15 years. The majority of computational models (starting with the McCulloch–Pitts neuron (McCulloch and Pitts, 1943)) assume a linear integration of inputs followed by a non-linear transfer function (Gazzaniga, 2004) (Figure 1A). Experimentally it is now well established that biological neurons are significantly more complex, particularly with respect to synaptic integration. Excitatory neurons, specifically pyramidal cells, have thousands of synapses (Spruston, 2008) and a complex dendritic structure (Figure 1B). Proximal synapses, those closest to the cell body, have a relatively large effect on the likelihood of a cell generating an action potential.

**[Figure 1 goes about here, please see end of manuscript]**

However, a majority of the synapses are distal, or far from the cell body. The activation of a single distal synapse (and its associated subthreshold potential) has almost no effect at the soma, but the coincident activation of a cluster of 8-20 spatially localized synapses generates a large regenerative NMDA spike (Spruston, 2008; Larkum et al., 2009; Antic et al., 2010). Synaptic activity must be spatially localized (typically within 20-300 microns) and tightly synchronized (typically within 1-5 milliseconds) to generate an NMDA spike (Major et al., 2013; Kleindienst et al., 2011; Takahashi et al., 2012). The spikes can then depolarize the cell for an extraordinary duration, sometimes for 50-200 milliseconds. Thus spatially localized dendritic segments act as coincidence detectors and provide a means for the many distal synapses to play a significant role in the cell's activity (Larkum et al., 2004; Smith et al., 2013). See (Antic et al., 2010; Major et al., 2013; Kastellakis et al., 2015) for detailed reviews of experimental findings and biophysical mechanisms. It is now generally understood that pyramidal cells contain a large number of non-linear active dendritic segments that independently modulate cell responses (Branco and Häusser, 2011; Major et al., 2013). The majority of information stored in a neuron may therefore be in the form of small independent synaptic clusters. A pyramidal neuron with ten thousand synapses could potentially have over one hundred such clusters.

Sparse representations are ubiquitous in neocortex, and the basic properties of active dendrites and pyramidal cells are thought to be consistent throughout neocortex (Spruston, 2008). It behooves us to ask whether there exist a common set of governing principles related to active dendrites and sparsity that are universal and independent of modality. A number of studies have explored sparse representations from a theoretical perspective. Early work on sparse coding (Olshausen and Field, 1997) suggests that sparse representations of the type found in V1 might satisfy information theoretic optimality criteria. Additional analysis (Kanerva, 1988; Rolls and Treves, 1990; Olshausen and Field, 2004; Nadal and Toulouse, 2009; Babadi and Sompolinsky, 2014) suggests that sparse representations may be particularly convenient for learning and memory systems. However none of these studies have considered models with active dendrites. With a few notable exceptions (Poirazi and Mel, 2001; Legenstein and Maass, 2011) theoretical studies have ignored active dendrites altogether. The studies that do model active dendrites have typically not explicitly incorporated sparse representations.

In this paper we develop a novel mathematical theory of sparse distributed representations and networks of neurons with active dendrites. We develop our analysis first from the perspective of a single dendritic segment and then from the perspective of a population of neurons. We propose that neurons form just a few synapses to a sparse subsample of active cells in activity patterns they need to recognize. In other words, a small set of spatially localized synapses on a dendritic branch can detect a prototypical pattern in a large population of active cells. We derive theoretical results that characterize the learning capacity and robustness of such representations in the context of neural tissue. The analysis demonstrates that individual neurons have the capacity to learn and classify a large number of patterns under extremely unreliable and noisy conditions. The results hold even when dendrites form synapses with a small subset of the cells in the pattern they want to recognize. Additionally, computer simulations of the equations provide some detailed numerical insights, such as natural bounds on the sparsity of representations, and the specific number of synapses required to initiate dendritic spikes. The results provide an explanation for the experimental findings and suggest that, through active dendrites, neurons can operate on sparse distributed representations in a highly robust and efficient manner.

## 2.    Materials and Methods

We assume an abstract neuron model with a collection of active dendrites. Our model departs from the traditional integrate and fire neurons used in artificial neural network models (Figure 1A). Instead we treat the neuron as a set of independent dendritic segments, each detecting one or more patterns of activity in some presynaptic area. This is consistent with the models discussed in (Poirazi et al., 2003; Larkum et al., 2009; Wu and Mel, 2009), a prototype of which is shown in Figure 1C.

Note that in real neurons synaptic input can be classified by where the dendrite segment is on the cell (proximal, distal basal, and apical) and where the afferents that connect to the dendrite segment originate. These dendritic zones are shown in Figure 1B and the location of the dendritic segment can affect somatic integration and spiking. In this paper we do not explicitly model the conversion of dendritic spikes into somatic spikes. This is an active area of research and there are several existing theories. (Poirazi et al., 2003) treat the neuron as a two-layer perceptron. (Jahnke et al., 2013; Breuer et al., 2014) have incorporated active dendrites into recurrent neural networks for synchrony and recalling precisely timed sequences. We have proposed a model that utilizes networks of neurons with active dendrites to form a powerful sequence memory mechanism (Hawkins and Ahmad, 2016).

Instead this paper focuses on the recognition capability of individual active dendritic segments, as shown in Figure 2, without regard to where the segment is on the neuron or the presynaptic source of synapses on the segment. In our analysis the segment is generic and agnostic – it simply receives some input and decides whether or not to initiate a dendritic spike. The overall recognition accuracy of a neuron in any of the above models will be bounded by the accuracy of its dendritic segments and the underlying representation. As such, an analysis of these two components can be used to provide insights into the overall capabilities of these models.

### 2.1.    Model Dendrite

Figure 2 illustrates our segment model and how a segment detects patterns. We model the instantaneous activity of presynaptic cells as being either on or off. The effect of an individual synapse is similarly binary. As shown in Figure 2A, a dendritic segment would typically be connected to a very small subset of all possible neurons in an input region. Although this paper focuses on a static analysis (i.e. we do not model learning), the existence of plasticity rules reliant on synapse clustering and NMDA spikes has been shown experimentally (Takahashi et al., 2012; Losonczy and Magee, 2006; Makino and Malinow, 2011; Makara et al., 2009). The specific subset of synapses within a segment will change over time as a result of structural plasticity (Chklovskii et al., 2004). Therefore the number of *potential connections* to a segment via plasticity is much larger than the number of actual connections. The actual connections are a subset of prototype activity patterns to be recognized.

**[Figure 2 goes about here, please see end of manuscript]**

Formally, we denote the number of potential connections as $n$. A dendritic segment is represented as a binary vector $\boldsymbol{D} = [b_0, \cdots, b_{n-1}]$ where a non-zero value $b_i$ indicates a synaptic connection to presynaptic cell $i$ and $s = |\boldsymbol{D}|$ indicates the number of synapses on that segment. Experimental findings suggest that typical numbers for $s$ are between 20 and 300 (Major et al., 2013). $n$, the number of potential connections is assumed to be much larger (numbering in the thousands) leading to very sparse $\boldsymbol{D}$.

Similarly, the binary vector $\boldsymbol{A}_t$ represents presynaptic activity (the activity of all potential connections) at time $t$. $\boldsymbol{A}_t$ is of dimensionality $n$ with $a_t = |\boldsymbol{A}_t|$ as the number of active cells. The activity is assumed to be sparse, i.e. $a_t \ll n$. Figure 2B illustrates such a scenario. When a threshold $\theta$ of the synapses on $s$ fire simultaneously, this is a "match" and causes a dendritic spike:

$$match(\boldsymbol{A}_t, \boldsymbol{D}) \equiv \boldsymbol{A}_t \cdot \boldsymbol{D} \geq \theta \qquad (1)$$

The binary dot product computes the overlap between the presynaptic area and the synapses stored on the segment; it is simply the number of bits that are "1" in the same locations in both vectors.

The use of very sparse vectors models the biology of active dendrites and is similar to the model in (Wu and Mel, 2009). It is a departure from most other computational models that typically assume full connectivity between a cell and its input area. An important property of our model is that activity in the presynaptic area that does not correspond to a synaptic connection on a segment has no impact on that segment. The rest of this paper explores mathematical properties of the model dendrite as it relates to pattern recognition.

## 3. Results

### 3.1. Accuracy of detecting a single pattern on a single segment

The synapses on segments represent a small subsample of the full patterns of activity in presynaptic areas. Given this, how reliably can a single segment detect the larger pattern? We assume that through learning the segment has created synapses to a subset of the active cells corresponding to a given pattern. The segment thus represents some prototypical pattern. We consider the problem of robustly detecting repeat occurrences of that pattern under various distortions. The degree to which a presynaptic pattern has to match the synapses is controlled by $\theta$, the NMDA spike threshold. Lower values of $\theta$ lead to more stability in recognition, i.e. the lower the value of $\theta$ with respect to $s$, the more deviation or noise the segment can tolerate. This comes at a cost since the lower the value of $\theta$ the more likely the segment would falsely detect a match to a different pattern. Thus there are inherent tradeoffs between the number of synapses per segment, the segment threshold, and the number of potential connections that affect overall noise robustness and the possibility of false matches. This is illustrated conceptually in Figure 3.

**[Figure 3 goes about here, please see end of manuscript]**

In order to quantify this effect, we introduce the notion of an "overlap set". Let $X$ be a binary SDR vector with $n$ components. The overlap set of $X$ with respect to $b$, $\Omega_X(n, a, b)$, is defined as the set of vectors of size $n$ with $a$ bits on that have exactly $b$ bits of overlap with $X$. The cardinality of this set, $|\Omega_X(n, a, b)|$, is the number of such vectors. Assuming $b \leq |X|$ and $b \leq a$, $|\Omega_X(n, a, b)|$ can be computed as the number of subsets of $X$ with $b$ bits ON, multiplied by the number of other patterns containing $n - |X|$ bits, of which $a - b$ bits are ON. In other words, this is:

$$|\Omega_X(n, a, b)| = \binom{|X|}{b} \times \binom{n - |X|}{a - b} \qquad (3)$$

We can now answer the following question: given a dendritic segment $D$ and random presynaptic activity pattern $A_t$, what is the probability of a match, $P(match(A_t, D))$? This is simply the number of possible matching vectors divided by the number of possible patterns:

$$P(match(A_t, D)) = \frac{\sum_{b=\theta}^{s} |\Omega_D(n, a_t, b)|}{\binom{n}{a_t}} \qquad (4)$$

It may be difficult to get an intuitive sense for this equation but as $n$, $s$ and $\theta$ increase, the denominator increases more than exponentially faster than the numerator. This means that the "grey area" representing the overall space in Figure 3 increases much faster than any of the "white areas" representing possible matches. In other words by increasing the parameters $n$ and $s$ it is easy to make the chance of false positives be extremely small.

Eq. (4) computes the probability of a false positive, but what about false negatives? If a pattern corresponding to one stored on a segment is corrupted with noise, it is possible that fewer than $\theta$ synapses will overlap. Assume a dendritic segment $D$ represents a subsample of some presynaptic activity pattern $A_t$ using $s$ synapses. Let $A_t^*$ represent a corrupted version of $A_t$ such that $v$ of the ON bits are now off. If $v$ is sufficiently small, i.e. $v \leq s - \theta$, the probability of a false negative is 0. As $v$ increases the probability of a false negative, i.e. $A_t^* \cdot D < \theta$, increases. We can compute the probability of such false negatives in a similar manner as above, by using overlap sets. The number of vectors $A_t^*$ that have exactly $b$ bits of overlap with $D$ is:

$$|\Omega_D(a_t, v, b)| = \binom{s}{b} \times \binom{a_t - s}{v - b} \qquad (5)$$

Thus the probability of a false negative is:

$$P(A_t^* \cdot D < \theta) = \frac{\sum_{b=s-\theta+1}^{s} |\Omega_D(a_t, v, b)|}{\binom{a_t}{v}} \qquad (6)$$

In general $a_t$, the number of cells active in $A_t$ will be greater than $s$, the number of synapses on a segment. $s$ will thus represent a (possibly very small) subsample of the actual pattern but the equations suggest it is still possible to get robust performance yielding a very efficient storage mechanism. To get a sense of the actual numbers for realistic scenarios, consider the following example. Suppose $n = 10,000$ and $s = 30$. If the overall sparsity is 3% (i.e. $a_t = 300$ active cells) the segment is subsampling a tenth of the bits in $A_t$. With a threshold $\theta = 12$, the probability of a false match is about one in $10^{12}$. In other words, even with 10% subsampling there is a negligible chance of a false positive with a random pattern. Even with a 20% corruption of $A_t$ (i.e. 60 bits, double the number of synapses on the segment), the probability of a false negative is smaller than one in $10^7$. In Section 4 we will show the range of robust parameters through simulations.

## 3.2.    Accuracy of a population of segments

Neurons have thousands of synapses arranged along numerous dendrite segments. It only takes 8 to 20 active synapses on a short section of dendrite to generate an NMDA spike (Major et al., 2013). Therefore a neuron has the potential to recognize hundreds of unique patterns. Further, in any cortical region, millions of neurons are each simultaneously trying to recognize hundreds of patterns. Will the robustness exhibited by a single dendrite segment be maintained in a region of neural tissue?

Consider the case of $M$ patterns represented by M independent dendritic segments (potentially on different cells), each with $s$ synapses and a threshold of $\theta$. In this scenario, each segment represents one pattern and it is a false positive if any of the segments falsely detect a different pattern.

Let $S$ be a set of $M$ vectors, $S = \{\boldsymbol{D}_0, \cdots, \boldsymbol{D}_{M-1}\}$ where each vector $\boldsymbol{D}_i$ represents a single dendritic segment. Given random presynaptic input $\boldsymbol{A}$, we classify it as belonging to this set as follows:

$$\boldsymbol{A} \in S \;\stackrel{\text{def}}{=}\; \exists_{\boldsymbol{D}_i \in S} match(\boldsymbol{D}_i, \boldsymbol{A}) = \; true \tag{7}$$

Given a new pattern $\boldsymbol{A}_t$, how reliably can we classify it? Here we assume the number of noise bits to be $\leq s_i - \theta$ where $s_i = |\boldsymbol{D}_i|$ denotes the number of synapses on $\boldsymbol{D}_i$. As such there are no false negatives, only false positives. The probability of a false positive (one or more matching segments) is:

$$P(\boldsymbol{A} \in S) = 1 - (1 - P(match(\boldsymbol{A}_t, \boldsymbol{D}_i)))^M \tag{8}$$

In practice the probability of an individual overlap is extremely small and it is difficult to compute without numerical issues. It is useful to use instead the following bound:

$$P(\boldsymbol{A} \in S) \leq M \cdot P(match(\boldsymbol{A}_t, \boldsymbol{D}_i)) \tag{9}$$

For sparse high dimensional vectors, with parameters in the ranges we are concerned with, Eq. (9) is in fact a very tight upper bound. To get a sense of the numbers and the overall capacity, consider the following example. Suppose n=10,000 and you have 3% sparsity ($a_t = 300$). By storing 30 synapses per segment using a segment threshold of 15, you can detect a million random SDR patterns with a false positive rate better than 1 in a billion!

This result points to a remarkable property of high dimensional sparse representations. You can convert patterns to a set of decorrelated high dimensional SDRs and simply store a small bit-wise subsample of each one. You can then classify a massive number of these patterns almost perfectly even in the presence of a large amount of noise and system unreliability. Thus a large collection of independent neurons, each with an independent set of segments, can robustly classify a very large number of patterns with a relatively small number of synapses.

### 3.3.    The union property

We have shown that a small number of synapses can reliably detect patterns in large populations of cells. The synapses recognizing a given pattern are typically spatially co-located on a dendritic segment (Kleindienst et al., 2011). However it is highly unlikely that synapses on dendritic segments are cleanly segregated into individual patterns. It is much more likely that synapses are somewhat mixed together. This is consistent with experimental results which show that a dendritic segment can contain hundreds of synapses yet a relatively small number of active synapses anywhere on the segment can trigger an NMDA spike (Major et al., 2008, 2013). In this section we consider what happens if the synapses to recognize multiple patterns are mixed together within a given dendritic segment. We show that our model dendrite can maintain robust recognition when multiple sets of synapses dedicated to recognizing different patterns are mixed together on a common dendritic segment.

It turns out that one of the properties of high dimensional sparse representations is their ability to reliably store a set of patterns within a single vector (Kanerva, 1988). We call this the "union property" as it involves creating a union (binary OR) of multiple patterns. Suppose we allow a single segment to contain $s$ synapses from $A_{t1}$ and an additional $s$ synapses from $A_{t2}$. The segment would spike if at least $\theta$ cells from $A_{t1}$ or $A_{t2}$ are active. By adding additional groups of $s$ synapses from other patterns, this segment will detect additional patterns. The vector representation of such a segment consists of the binary OR of the $s$ synapses from each pattern. We say that this segment now represents a union of patterns and any combination of $\theta$ active synapses from this union will cause a spike.

The advantage of a union is that a fixed SDR element (such as that represented by a single segment) can be used to recognize a varying number of patterns. There will never be a false negative: the segment will reliably fire in the presence of any of the stored activity patterns with up to $s - \theta$ bits of noise. The downside of course is that there is now a larger potential for false positives. It is possible for the segment to spike due to a mixture of active cells, say half from $A_{t1}$ and half from $A_{t2}$. As you add more patterns to the union there are an increasing number of mismatch possibilities and the segment is increasingly likely to spike for random patterns. Although forming a union introduces a potentially significant source of error, we will show that with high dimensional vectors pattern recognition can still be performed very reliably using such a union representation.

Formally, the mechanics of unions are simple. To "store" a set of $M$ sparse patterns we simply take the Boolean OR of all of them to create a new binary vector $\mathbf{X}$ (Figure 4).

$$X = \bigcup_{i=0}^{M-1} x_i \tag{10}$$

Some of the bits in $x_i$ may overlap so $X$ is now a binary vector such that $|X| \leq \sum_{i=0}^{M-1} |x_i|$. To check if a new pattern $\mathbf{y}$ is a member of the set, we compute the match as in Eq. (1).

How reliable is the classification operation? We first consider the case of exact matches, i.e. $\theta = |y|$. For simplicity we also assume all vectors in $X$ have the same number of ON bits. Note that if $y = x_i$ for some $i$, the match operation will always be successful. However for other vectors there is a chance of a false positive match due to mix and match errors. A false positive with a new random pattern $y$ occurs if all of the bits in $y$ overlap with $X$. When $M = 1$, the probability that any given bit is 0 is $1 - q$, where $q = \frac{|x_0|}{n}$. As M grows, the probability that a given bit is still 0 is:

$$p_0 = (1 - q)^M \tag{11}$$

The probability that a given bit in **X** is ON is therefore $(1 - p_o)$. The probability of a false positive match, i.e. that all the bits in **y** are ON is therefore:

$$P(match(\boldsymbol{X}, \boldsymbol{y})) = (1 - p_0)^{|\boldsymbol{y}|} = (1 - (1 - q)^M)^{|\boldsymbol{y}|} \tag{12}$$

This calculation is equivalent to the probability of false positives in Bloom filters (Bloom, 1970) and very similar to the analysis of sparse memory in Willshaw networks (Nadal and Toulouse, 2009). We can now plug this back into Eq. (4) to compute the probability of error for inexact matches. After M union operations, the expected number of ON bits in X is:

$$E[|\boldsymbol{X}|] = (1 - p_0)n \tag{13}$$

The expected size of the overlap set is:

$$E[|\Omega_X(n, a, b)|] = \binom{(1 - p_0)n}{b} \times \binom{n - (1 - p_0)n}{a - b} \tag{14}$$

We can now calculate how accurately a single dendritic segment can represent a mix of different patterns. Suppose a segment **D** represents a union of M different patterns. The expected number of synapses on this segment $s$ is $E[s] = (1 - p_0)n$. Given random presynaptic activity $\boldsymbol{A}_t$, the expected probability of a false match is therefore:

$$E[P(match(\boldsymbol{D}, \boldsymbol{A}_t))] = \frac{\sum_{b=\theta}^{s} E[|\Omega_D(n, a, b)|]}{\binom{n}{a}} \tag{15}$$

The equation is complex but the numbers can be illuminating. As an example, suppose the population size $n = 20000$, with $a = 100$ cells active at a time. Suppose each pattern on a segment is represented by 25 synapses and $\theta = 15$. If you union together $M = 10$ patterns, on average you would get fewer than 250 synapses on the segment. In this scenario the false positive rate is less than one in $10^{11}$. To gain an intuition for this, it is useful to think about the expected number of ON bits in the SDR, i.e. Eq. (13). With $M = 10$, 98.75% of the bits in the vector representing this segment are zero. When you match against an additional vector, there is a very high chance it will have most of its bits among this 98.75%, and hence it won't lead to a false positive. Only vectors that have at least 15 of their bits ON among the 1.5% will cause a false positive.

The net impact of the union property is that individual dendritic segments can be sloppy and reliable at the same time. There are limits, but such segments can mix together a number of independent patterns with virtually no chance of false positive errors. The equations do suggest a higher threshold when multiple patterns are mixed together in order to maintain a given error rate. This is consistent with the finding that a larger number of active synapses are required to initiate an NMDA spike when the synapses are spread out over a dendrite's length (Major et al., 2013; McBride et al., 2008). The union property provides some theoretical foundation for the proposal that a longer stretch of dendrite can reliably function as a more flexible decision making unit detecting multiple patterns (Major et al., 2008).

## 4.    Simulation Results

Due to the various factorials and exponentials involved in the above equations it is sometimes difficult to develop a solid intuitive understanding of the various scaling properties. The range of parameters leading to robust recognition may be unclear, and it may also be unclear how these results translate to real neurons and experimental data. In this section we describe a number of simulation results[2]. Our goal is to develop intuitions for the reader, and to demonstrate the applicability of the results.

## 4.1.    Numerical experiments with an artificial dendrite

We first report results from numerical experiments that randomly sampled large numbers of vectors and explicitly calculated matches using a simple threshold dendritic model. These simulations are used to verify the equations for false positives (Eq. (4)) and false negatives (Eq. (6)). We also used this method to simulate the effects of noise and estimate the associated probability of false negatives.

Figure 4A shows numerically simulated and predicted false positive rates. To compute false positive rates we first create a prototype dendrite that samples $s$ synapses from a randomly generated presynaptic vector with $a$ bits out of $n$. We then randomly generate a second sparse vector and test whether the two vectors match. The entire process is then repeated 100 million times for each tested combination of $s$, $a$, and $n$. (Due to the extremely low error rates, in order to obtain reliable results these experiments require a large sampling of the space.) As can be seen from the chart, there is virtually no difference between theoretical and experimentally observed false positive rates.

**[Figure 4 goes about here, please see end of manuscript]**

Figure 4B shows a similar experiment exploring the probability of false negatives. To compute false negative rates we first create a prototype dendrite that samples $s$ synapses from a randomly generated presynaptic vector $\boldsymbol{A}$ with $a$ bits out of $n$. We then randomly generate a noisy version $\boldsymbol{A}^*$ with up to v ON bits swapped with other bits. We then tested whether the subsampled dendrite matched the noisy version $\boldsymbol{A}^*$.  As before we repeated the entire process a large number of times (10 million for 4B). The theoretical and observed false negative rates are virtually identical.

It is impractical to accurately calculate errors through random sampling for realistic numbers of synapses and neurons (an error rate of 1 in $10^{12}$ requires a sampling size of at least $10^{13}$). Nevertheless, these experiments are sufficient to show that the theoretical predictions from Eqs. (4) and (6) match simulation results closely. In subsequent experiments we calculate the equations directly using high precision math libraries. This allows us to explore a larger part of the parameter space and draw conclusions for numbers that are closer to those in biology.

## 4.2.    Subsampling and the effects of sparsity and dimensionality

We ran a number of simulations to illustrate some of the properties of dendritic matching (i.e. Eq. (4)). Figure 5A shows the effect of the underlying dimensionality of the representation space, i.e. the population of cells. We plot the drop in error rates as you increase the population size, $n$, while maintaining a fixed sparsity level and a fixed number of synapses on a segment. The error drops rapidly (faster than exponentially) as $n$ increases, becoming essentially 0 once $n > 2000$. Note that it is not possible to get robust recognition with a dense representation, as shown by the dashed line representing a 50% activity level. Thus both sparsity and high dimensionality are required to achieve robust recognition with a small number of synapses.

---

**[Figure 5 goes about here, please see end of manuscript]**

Figure 5B examines the effect of changing the number of synapses on a dendritic segment. It illustrates how segments can store a tiny subsample from a large population and still robustly recognize complex patterns. If the population of cells increases beyond a few thousand, and the overall activity is sparse, it is possible to achieve reliable recognition with a small sample of each pattern. The chart shows that the error rate decreases exponentially with the number of synapses. For many situations a subsample of 20 to 25 synapses leads to an error rate better than $10^{-10}$. Note that since $\theta = \frac{s}{2}$, this includes a noise level up to 50%. This helps explain how even a small number of synapses on a segment are sufficient for robust recognition performance. The dashed line shows that denser representations lead to high error rates. Although not shown, small $a$ and $n$ (e.g. $a = 32$ and $n = 128$) also lead to high error. In other words, in order to achieve accurate recognition with a small subsample, you need both sparsity and a sufficiently high dimensionality.

Eq. (4) was calculated directly to obtain the above results. In order to verify the accuracy of the equations we also ran some numerical simulations by randomly sampling large numbers of vectors and explicitly calculating matches using a simulated dendritic model.

### 4.3.    Estimating the optimal spike threshold

The mathematics behind sparse representations can help provide insight into key experimental results. In Figure 6 we show that the equations can be used to explore the effect of different dendritic spike thresholds and suggest an optimal range. Figure 6 shows the median probability of error as a function of the synaptic threshold for a dendritic spike. Each point on the graph holds the threshold $\theta$ fixed and represents the median probability of error computed over a large range of all other parameters: $n$ (the number of potential synapses), presynaptic activity $a$, and $s$ the number of synapses on each segment. We systematically varied $n$ from 10,000 to 200,000, presynaptic activity from 0.5% to 3% of $n$, and $s$ from 20 to 50. The shaded area of the chart is the region corresponding to low error (for illustration purposes we show the range of thresholds that lead to an error of 1 in a billion or lower).

Figure 6 demonstrates that a NMDA spike threshold of 9 and higher leads to low error rates under a very wide range of assumptions. A spike threshold beyond 15 or 20 leads to diminishing returns. Beyond 20 the error rates are so low that the extra metabolic cost of forming additional synapses due to higher thresholds is not justified. Thus the equations predict that the optimal dendritic spiking threshold is between 9 and 20. This lines up well with experimental results, which show actual spiking thresholds to be between 8 and 20 (Major et al., 2013; Branco and Häusser, 2011).

**[Figure 6 goes about here, please see end of manuscript]**

### 4.4.    The effect of unions on a segment

Figure 7 shows two simulations that demonstrate the effect of the union property on a dendritic segment. As discussed earlier, a segment can contain synapses from a mixture of independent patterns and initiate an NMDA spike if any of them are detected. Figure 7A shows how the expected number of synapses on a segment scales with the number of patterns (i.e. Eq. (13)). Experimental results show that a single dendritic segment can contain anywhere from 100 and 400 synapses (Major et al., 2013). The graph suggests that this translates to between 4 and 16 independent patterns, dependent on the number of synapses used to represent each pattern. Figure 7B illustrates the probability of mix and match errors for segments that represent such a union of patterns. Here $n$, the size of the presynaptic population, and the overall sparsity levels are critical factors. As the graph shows, a larger presynaptic population implies a low chance of mix and match errors. A small presynaptic population of 1000 leads to relatively high error rates, but a population of 20,000 can lead to extremely low error rates even with 10 patterns stored. Overall these graphs show that it is possible for a single dendritic segment to sloppily store a mixture of multiple patterns, yet maintain remarkably high accuracy for detecting each pattern.

**[Figure 7 goes about here, please see end of manuscript]**

Note that the error rates plotted in Figures 4-7 are for individual segments. However, the correct classification of a pattern is always performed by a population of neurons. Populations of active cells can contain a significant number of incorrect false positives activations without error in classification of the entire population. Therefore, even if individual neurons operate in a region of unacceptably high false positives, the population accuracy will be substantially lower.

## 5.        Discussion

Sparse distributed representations are ubiquitous in neocortex. In this paper we have proposed a formal mathematical model for sparse representations in neocortex based on properties of active dendrites. Our model and the core operations of overlap and match are inspired by experimental findings on active dendritic processing and NMDA spikes in pyramidal neurons. We derived a number of scaling laws demonstrating that systems based on these principles can achieve extreme robustness to noise and faults in the system. Our simulation results provide detailed insights into various parameter regimes and show that both sparsity and high dimensionality are required for maximum accuracies. In addition we show that the equations can be used to predict experimental results, such as the optimal spiking thresholds for active dendrites.

Our work is directly related to the theoretical work of (Kanerva, 1988; Rolls and Treves, 1990; Olshausen and Field, 2004) as well as the work on active dendrites (Poirazi and Mel, 2001). (Rolls and Treves, 1990) were perhaps the first to make the connection between the improved robustness to noise afforded by sparse representations and the role of non-linear dendrites and NMDA receptors. A difference with these papers is that they use a typical Euclidean distance norm and a weighted linear sum with scalar vectors instead of our overlap metric with binary vectors. (Poirazi and Mel, 2001) showed increased capacity and lower error rates in neurons with active dendrites compared to linear neurons with a similar number of synapses. They did not explicitly compute the impact of increased dimensionality and increased sparseness. Specifically, their experimental simulations were limited to a relatively low dimensionality of 400, a sub-optimal range according to Eq. (4). (The number of synapses in their simulations, 25, does fall in a good range as demonstrated by Figure 5B.) Our analysis thus suggests that the error rates reported in their simulations would be significantly improved with dimensionalities and sparsities in the ranges indicated by Eq. (4) and Figure 5A.

(Babadi and Sompolinsky, 2014) have also developed a theoretical model of sparsity that is close in spirit to our results. They explicitly test the hypothesis that expanded sparse representations of an input space lead to improved reliability and noise robustness (they observe that primary sensory areas typically undergo a 25:1 expansion of the axons entering the area). They too generally conclude that sparsity in high dimensional spaces is desirable. However they do not explicitly model non-linear active dendrites (they use a linear readout) and do not analyze the highly sparse connectivity consistent with the literature on active dendrites. The error rates and noise tolerances suggested by our results are orders of magnitude lower than the readout errors reported in their paper. The numerical results in our paper also more closely match known experimental data on active dendrites. The impact of sparse encodings in the context of associative networks has been studied by (Rolls and Treves, 1990; Nadal and Toulouse, 2009). (Sommer and Palm, 1999; Knoblauch et al., 2010) have also studied the impact of sparse representations and various forms of Hebbian plasticity on associative memories. Their analysis has focused on pattern completion and does not explicitly consider recognition and the benefits of subsampling. Despite the differences noted, taken together the above papers provide general support for the power of sparse high dimensional representations, and the insights to be gained from mathematical modeling and simulations of sparsity.

In this paper we have discussed neurons with active dendrites, and their error rates in pattern detection. We have not discussed the functional implication of these matches. What does the neuron do once a dendritic match occurs? The functional benefits of active dendrites are a topic of active research, and there have been several theories proposed in the literature. These theories include translation invariance (Mel et al., 1998), the efficient propagation of neural activity (Polsky et al., 2009), facilitating top-down prediction (Larkum, 2013), sequence storage (Losonczy et al., 2008; Branco et al., 2010), and gain control (Larkum et al., 2004). In the context of recurrent networks (Jahnke et al., 2013; Breuer et al., 2014) have shown that active and non-linear dendrites can facilitate the propagation of synchrony and the replay of precisely timed sequences. Finally, elsewhere we have described a detailed theory and working implementation that shows how active dendrites and networks of pyramidal cells lead to a sophisticated and practical sequence memory algorithm (Hawkins and Ahmad, 2016). The overall capabilities of all of these models are bounded by the robustness of the underlying dendritic representations. As such, the mathematical framework proposed in this paper can be used to provide insights into the overall power of these models and suggest optimal ranges for dendritic parameters.

The equations in this paper assume random and decorrelated neural activity. The distribution of individual spiking neurons and neural correlation in neocortex is a topic of some debate (Cohen and Kohn, 2011). A number of papers have suggested that one of the outcomes of neural plasticity and inhibition is to decorrelate the inputs. For example, Hebbian-style learning plus inhibition leads to individual neurons that represent successive principal components of the input space (Oja, 1982). (Barlow and Földiák, 1989) discuss the theoretical benefits of decorrelated responses in cortex. Indeed in-vivo measurements of neural activity imply correlation is low, even for neurons with highly overlapping receptive fields (Smith and Häusser, 2010; Ecker et al., 2010). (Vinje and Gallant, 2000) show that with natural image stimuli, the selectivity and sparseness of individual V1 neurons increases *in vivo*, and decorrelates the responses of neuron pairs. (Simoncelli and Olshausen, 2001) reviews additional experimental evidence demonstrating that correlated visual inputs are decorrelated as early as V1. In addition it is our belief that, due to the underlying robustness of sparse representations, it is not necessary to have completely uncorrelated random distributions to obtain high fault tolerance. It is clear that increased correlation will lead to higher than random probability of overlap between vectors. Eq. (6) and Figure 4 analyze this condition and suggest that a significant amount of overlap can be tolerated without impacting performance. Our analysis has focused on uniformly random distributions; extending them to other distributions is an interesting topic for future work.

The equations in this paper assume binary synapses and binary cell activations, and ignore the possibility of scalar weights or outputs. The question of binary synapses is a topic that is heavily debated in literature. For example some researchers have argued that synapses are inherently binary (Petersen et al., 1998), while others have argued the opposite (Enoki et al., 2009). (Amit and Fusi, 1994; Fusi and Abbott, 2007) have studied the memory capacity of bounded synapses. Interestingly, in some of their studies (Poirazi and Mel, 2001) found that binary receptive fields led to the best memory capacity, and 4-level synapses were optimal in others. For the purposes of this paper we simply note that the robustness demonstrated in this paper relies on sparseness, but does not rely on binary values. Synapses and cell outputs represented by higher resolution vectors are a superset of binary vectors and thus in theory can only add to the power demonstrated here. Consider the case where cell outputs (and synapses) can each take on $m$ different values. If every state can be equally discriminated, it would be possible to create an identical system using binary vectors of size $n \log_2 m$ and directly apply the equations in this paper. This would provide bounds although their tightness could potentially be improved. A comprehensive study of the scaling properties of multi-valued or continuous SDRs is beyond the scope of this paper, but represents an interesting direction for future research.

Individual neurons and synapses are inherently unreliable (Faisal et al., 2008), yet the overall system works extremely well. The results in this paper shed light on how cortical processing can be incredibly robust and fault tolerant as long as the underlying representation is sparse and high dimensional. The mathematical model described here is not specific to any sensory modality or cortical area. As such the properties should be ubiquitous for pyramidal cells everywhere. It is our hope that over time, a complete theoretical understanding of the learning and scaling properties of neocortical representations can be developed.

## Acknowledgments

We thank Yuwei Cui and Scott Purdy as well as numerous Numenta employees for their comments and feedback during the preparation of this article. We also thank the reviewers for their excellent feedback and help improving the clarity of this manuscript.

## References

Amit, D. J., and Fusi, S. (1994). Learning in Neural Networks with Material Synapses. *Neural Comput.* 6, 957–982. doi:10.1162/neco.1994.6.5.957.

Antic, S. D., Zhou, W. L., Moore, A. R., Short, S. M., and Ikonomu, K. D. (2010). The decade of the dendritic NMDA spike. *J. Neurosci. Res.* 88, 2991–3001. doi:10.1002/jnr.22444.

Babadi, B., and Sompolinsky, H. (2014). Sparseness and Expansion in Sensory Representations. *Neuron* 83, 1213–1226. doi:10.1016/j.neuron.2014.07.035.

Barlow, H. B., and Földiák, P. (1989). Adaptation and decorrelation in the cortex. *Comput. Neuron*, 54–72.

Barth, A. L., and Poulet, J. F. a (2012). Experimental evidence for sparse firing in the neocortex. *Trends Neurosci.* 35, 345–355. doi:10.1016/j.tins.2012.03.008.

Bloom, B. H. (1970). Space/time trade-offs in hash coding with allowable errors. *Commun. ACM* 13, 422–426. doi:10.1145/362686.362692.

Branco, T., Clark, B. A., and Häusser, M. (2010). Dendritic discrimination of temporal input sequences in cortical neurons. *Science* 329, 1671–1675. doi:10.1126/science.1189664.

Branco, T., and Häusser, M. (2011). Synaptic integration gradients in single cortical pyramidal cell dendrites. *Neuron* 69, 885–92. doi:10.1016/j.neuron.2011.02.006.

Breuer, D., Timme, M., and Memmesheimer, R.-M. (2014). Statistical Physics of Neural Systems with Nonadditive Dendritic Coupling. *Phys. Rev. X* 4, 011053. doi:10.1103/PhysRevX.4.011053.

Chklovskii, D. B., Mel, B. W., and Svoboda, K. (2004). Cortical rewiring and information storage. *Nature* 431, 782–8. doi:10.1038/nature03012.

Cohen, M. R., and Kohn, A. (2011). Measuring and interpreting neuronal correlations. *Nat. Neurosci.* 14, 811–819. doi:10.1038/nn.2842.

Crochet, S., Poulet, J. F. a, Kremer, Y., and Petersen, C. C. H. (2011). Synaptic mechanisms underlying sparse coding of active touch. *Neuron* 69, 1160–1175. doi:10.1016/j.neuron.2011.02.022.

Ecker, A. S., Berens, P., Keliris, G. A., Bethge, M., Logothetis, N. K., and Tolias, A. S. (2010). Decorrelated neuronal firing in cortical microcircuits. *Science* 327, 584–587. doi:10.1126/science.1179867.

Enoki, R., Hu, Y. ling, Hamilton, D., and Fine, A. (2009). Expression of Long-Term Plasticity at Individual Synapses in Hippocampus Is Graded, Bidirectional, and Mainly Presynaptic: Optical Quantal Analysis. *Neuron* 62, 242–253. doi:10.1016/j.neuron.2009.02.026.

Faisal, A. A., Selen, L. P. J., and Wolpert, D. M. (2008). Noise in the nervous system. *Nat. Rev. Neurosci.* 9, 292–303. doi:10.1038/nrn2258.

Fusi, S., and Abbott, L. F. (2007). Limits on the memory storage capacity of bounded synapses. *Nat. Neurosci.* 10, 485–493. doi:10.1038/nn1859.

Gazzaniga, M. S. [Ed] (2004). *The cognitive neurosciences (3rd ed.).* doi:10.1136/bmj.312.7024.193.

Graziano, M. S. A., and Aflalo, T. N. (2007). Mapping behavioral repertoire onto the cortex. *Neuron* 56, 239–251. doi:10.1016/j.neuron.2007.09.013.

Graziano, M. S. A., Taylor, C. S. R., and Moore, T. (2002). Complex movements evoked by microstimulation of precentral cortex. *Neuron* 34, 841–851. doi:10.1016/S0896-6273(02)00698-0.

Hawkins, J., and Ahmad, S. (2016). Why Neurons Have Thousands of Synapses, a Theory of Sequence Memory in Neocortex. *Front. Neural Circuits* 10. doi:10.3389/fncir.2016.00023.

Hromádka, T., DeWeese, M. R., and Zador, A. M. (2008). Sparse representation of sounds in the unanesthetized auditory cortex. *PLoS Biol.* 6, 0124–0137. doi:10.1371/journal.pbio.0060016.

Jahnke, S., Memmesheimer, R.-M., and Timme, M. (2013). Propagating synchrony in feed-forward networks. *Front. Comput. Neurosci.* 7, 153. doi:10.3389/fncom.2013.00153.

Kanerva, P. (1988). *Sparse Distributed Memory*. Cambridge, MA: The MIT Press.

Kastellakis, G., Cai, D. J., Mednick, S. C., Silva, A. J., and Poirazi, P. (2015). Synaptic clustering within dendrites: An emerging theory of memory formation. *Prog. Neurobiol.* 126, 19–35. doi:10.1016/j.pneurobio.2014.12.002.

Kiani, R., Esteky, H., Mirpour, K., and Tanaka, K. (2007). Object category structure in response patterns of neuronal population in monkey inferior temporal cortex. *J. Neurophysiol.* 97, 4296–4309. doi:10.1152/jn.00024.2007.

Kleindienst, T., Winnubst, J., Roth-Alpermann, C., Bonhoeffer, T., and Lohmann, C. (2011).

Activity-dependent clustering of functional synaptic inputs on developing hippocampal dendrites. *Neuron* 72, 1012–1024. doi:10.1016/j.neuron.2011.10.015.

Knoblauch, A., Palm, G., and Sommer, F. T. (2010). Memory capacities for synaptic and structural plasticity. *Neural Comput.* 22, 289–341. doi:10.1162/neco.2009.08-07-588.

Larkum, M. (2013). A cellular mechanism for cortical associations: an organizing principle for the cerebral cortex. *Trends Neurosci.* 36, 141–51. doi:10.1016/j.tins.2012.11.006.

Larkum, M. E., Nevian, T., Sandler, M., Polsky, A., and Schiller, J. (2009). Synaptic integration in tuft dendrites of layer 5 pyramidal neurons: a new unifying principle. *Science* 325, 756–760. doi:10.1126/science.1171958.

Larkum, M. E., Senn, W., and Lüscher, H. R. (2004). Top-down dendritic input increases the gain of layer 5 pyramidal neurons. *Cereb. Cortex* 14, 1059–1070. doi:10.1093/cercor/bhh065.

Legenstein, R., and Maass, W. (2011). Branch-specific plasticity enables self-organization of nonlinear computation in single neurons. *J. Neurosci.* 31, 10787–802. doi:10.1523/JNEUROSCI.5684-10.2011.

Losonczy, A., and Magee, J. C. (2006). Integrative properties of radial oblique dendrites in hippocampal CA1 pyramidal neurons. *Neuron* 50, 291–307. doi:10.1016/j.neuron.2006.03.016.

Losonczy, A., Makara, J. K., and Magee, J. C. (2008). Compartmentalized dendritic plasticity and input feature storage in neurons. *Nature* 452, 436–41. doi:10.1038/nature06725.

Major, G., Larkum, M. E., and Schiller, J. (2013). Active properties of neocortical pyramidal neuron dendrites. *Annu. Rev. Neurosci.* 36, 1–24. doi:10.1146/annurev-neuro-062111-150343.

Major, G., Polsky, A., Denk, W., Schiller, J., and Tank, D. W. (2008). Spatiotemporally graded NMDA spike/plateau potentials in basal dendrites of neocortical pyramidal neurons. *J. Neurophysiol.* 99, 2584–2601. doi:10.1152/jn.00011.2008.

Makara, J. K., Losonczy, A., Wen, Q., and Magee, J. C. (2009). Experience-dependent compartmentalized dendritic plasticity in rat hippocampal CA1 pyramidal neurons. *Nat. Neurosci.* 12, 1485–1487. doi:10.1038/nn.2428.

Makino, H., and Malinow, R. (2011). Compartmentalized versus global synaptic plasticity on dendrites controlled by experience. *Neuron* 72, 1001–1011. doi:10.1016/j.neuron.2011.09.036.

McBride, T. J., Rodriguez-Contreras, A., Trinh, A., Bailey, R., and DeBello, W. M. (2008). Learning Drives Differential Clustering of Axodendritic Contacts in the Barn Owl Auditory System. *J. Neurosci.* 28, 6960–6973. doi:10.1523/JNEUROSCI.1352-08.2008.

McCulloch, W. S., and Pitts, W. H. (1943). A logical calculus of ideas imminent in nervous activity. *Bull. Math. Biophys.* 5, 115–133. doi:10.1007/BF02478259.

Mel, B. W., Ruderman, D. L., and Archie, K. A. (1998). Translation-invariant orientation tuning in visual "complex" cells could derive from intradendritic computations. *J. Neurosci.* 18, 4325–4334.

Nadal, J.-P., and Toulouse, G. (2009). Information storage in sparsely coded memory nets. *Netw. Comput. Neural Syst.*

Oja, E. (1982). A simplified neuron model as a principal component analyzer. *J. Math. Biol.* 15, 267–273. doi:10.1007/BF00275687.

Olshausen, B. A., and Field, D. J. (2004). Sparse coding of sensory inputs. *Curr. Opin. Neurobiol.* 14, 481–487. doi:10.1016/j.conb.2004.07.007.

Olshausen, B. A., and Field, D. J. (1997). Sparse coding with an overcomplete basis set: A strategy employed by V1? *Vision Res.* 37, 3311–3325. doi:10.1016/S0042-6989(97)00169-7.

Petersen, C. C., Malenka, R. C., Nicoll, R. A., and Hopfield, J. J. (1998). All-or-none potentiation at CA3-CA1 synapses. *Proc. Natl. Acad. Sci. U. S. A.* 95, 4732–4737. doi:10.1073/pnas.95.8.4732.

Poirazi, P., Brannon, T., and Mel, B. W. (2003). Pyramidal neuron as two-layer neural network. *Neuron* 37, 989–99. Available at: http://www.ncbi.nlm.nih.gov/pubmed/12670427 [Accessed April 14, 2015].

Poirazi, P., and Mel, B. W. (2001). Impact of active dendrites and structural plasticity on the memory capacity of neural tissue. *Neuron* 29, 779–796. doi:10.1016/S0896-6273(01)00252-5.

Polsky, A., Mel, B., and Schiller, J. (2009). Encoding and decoding bursts by NMDA spikes in basal dendrites of layer 5 pyramidal neurons. *J. Neurosci.* 29, 11891–11903. doi:10.1523/JNEUROSCI.5250-08.2009.

Rolls, E., and Treves, A. (1990). The relative advantages of sparse versus distributed encoding for associative neuronal networks in the brain. *Netw. Comput. neural Syst.* Available at: http://www.tandfonline.com/doi/abs/10.1088/0954-898X_1_4_002 [Accessed April 12, 2016].

Simoncelli, E. P., and Olshausen, B. a (2001). Natural image statistics and neural representation. *Annu. Rev. Neurosci.* 24, 1193–1216. doi:10.1146/annurev.neuro.24.1.1193.

Smith, S. L., and Häusser, M. (2010). Parallel processing of visual space by neighboring neurons in mouse visual cortex. *Nat. Neurosci.* 13, 1144–1149. doi:10.1038/nn.2620.

Smith, S. L., Smith, I. T., Branco, T., and Häusser, M. (2013). Dendritic spikes enhance stimulus selectivity in cortical neurons in vivo. *Nature* 503, 115–20. doi:10.1038/nature12600.

Sommer, F. T., and Palm, G. (1999). Improved bidirectional retrieval of sparse patterns stored by Hebbian learning. *Neural Networks* 12, 281–297. doi:10.1016/S0893-6080(98)00125-7.

Spruston, N. (2008). Pyramidal neurons: dendritic structure and synaptic integration. *Nat. Rev. Neurosci.* 9, 206–221. doi:10.1038/nrn2286.

Takahashi, N., Kitamura, K., Matsuo, N., Mayford, M., Kano, M., Matsuki, N., and Ikegaya, Y. (2012). Locally Synchronized Synaptic Inputs. *Science (80-. ).* 335, 353–356. doi:10.1126/science.1210362.

Vinje, W. E., and Gallant, J. L. (2000). Sparse coding and decorrelation in primary visual cortex during natural vision. *Science (80-. ).* 287, 1273–6. doi:10.1126/science.287.5456.1273.

Weliky, M., Fiser, J., Hunt, R. H., and Wagner, D. N. (2003). Coding of natural scenes in primary visual cortex. *Neuron* 37, 703–718. doi:10.1016/S0896-6273(03)00022-9.

Wu, X. E., and Mel, B. W. (2009). Capacity-enhancing synaptic learning rules in a medial temporal lobe online learning model. *Neuron* 62, 31–41. doi:10.1016/j.neuron.2009.02.021.

# Figures

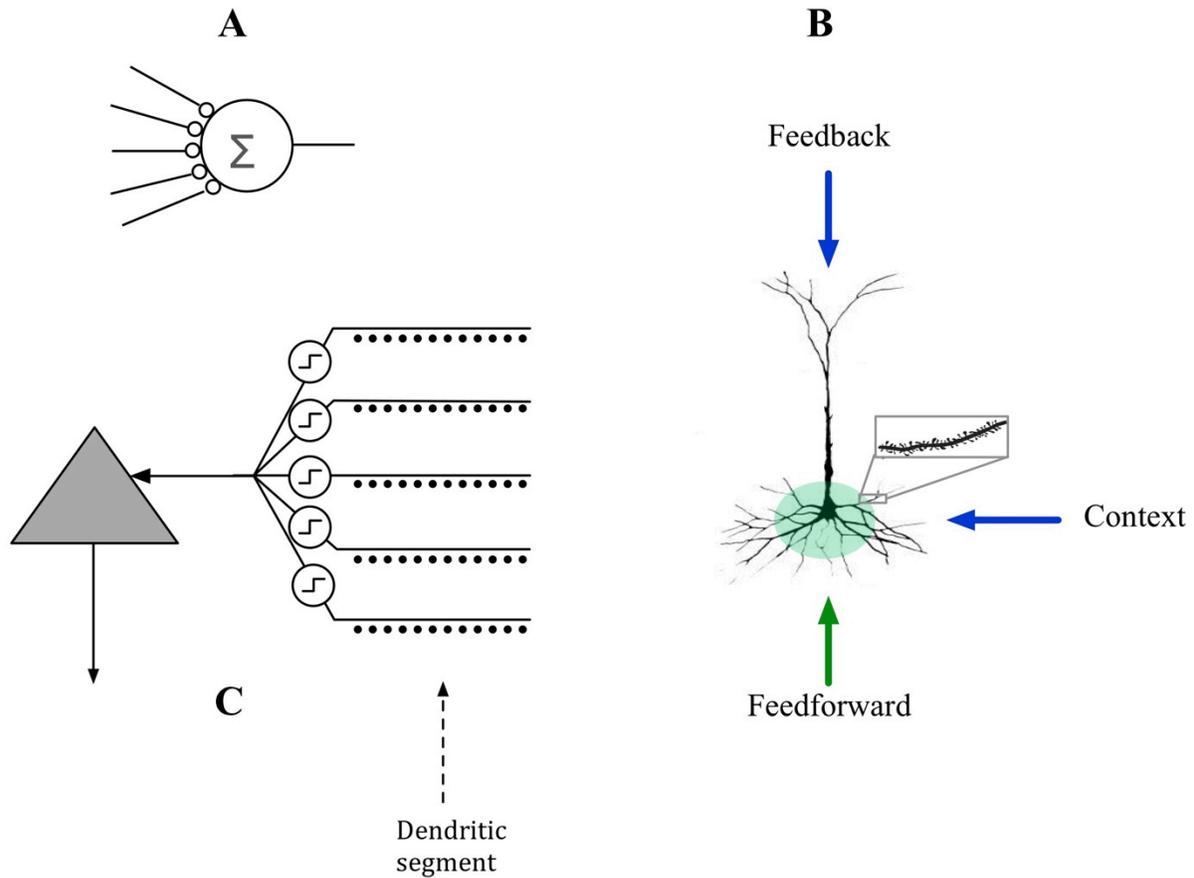

**A**

**B**

Feedback

Context

Feedforward

**C**

Dendritic
segment

# Figure 1

**Figure 1. A.** A point neuron used in most artificial neural networks. The output of the point neuron is a function of the sum of all the inputs; active dendritic properties are ignored. **B.** A neocortical pyramidal neuron has thousands of excitatory synapses that are located on dendrites. Synapses proximal to the cell body (green area) comprise <10% of a cell's inputs, receive feedforward input, and define the basic receptive field response of the cell. Synapses on the basal distal dendrites typically receive contextual input from nearby cells. Apical distal synapses typically receive feedback inputs. The activation of a single distal synapse often has no measurable effect at the soma. However, the activation of a small number of synapses in close spatial and temporal proximity on a basal distal dendrite can cause an NMDA spike and significant depolarization at the soma. **C.** A prototypical neuron composed of an array of active dendritic segments (only five shown). Each dendritic segment contains a number of synapses and is associated with a spiking threshold. If the instantaneous number of active synapses on a segment reaches threshold, the segment initiates a spike, thus acting as a coincidence detector.

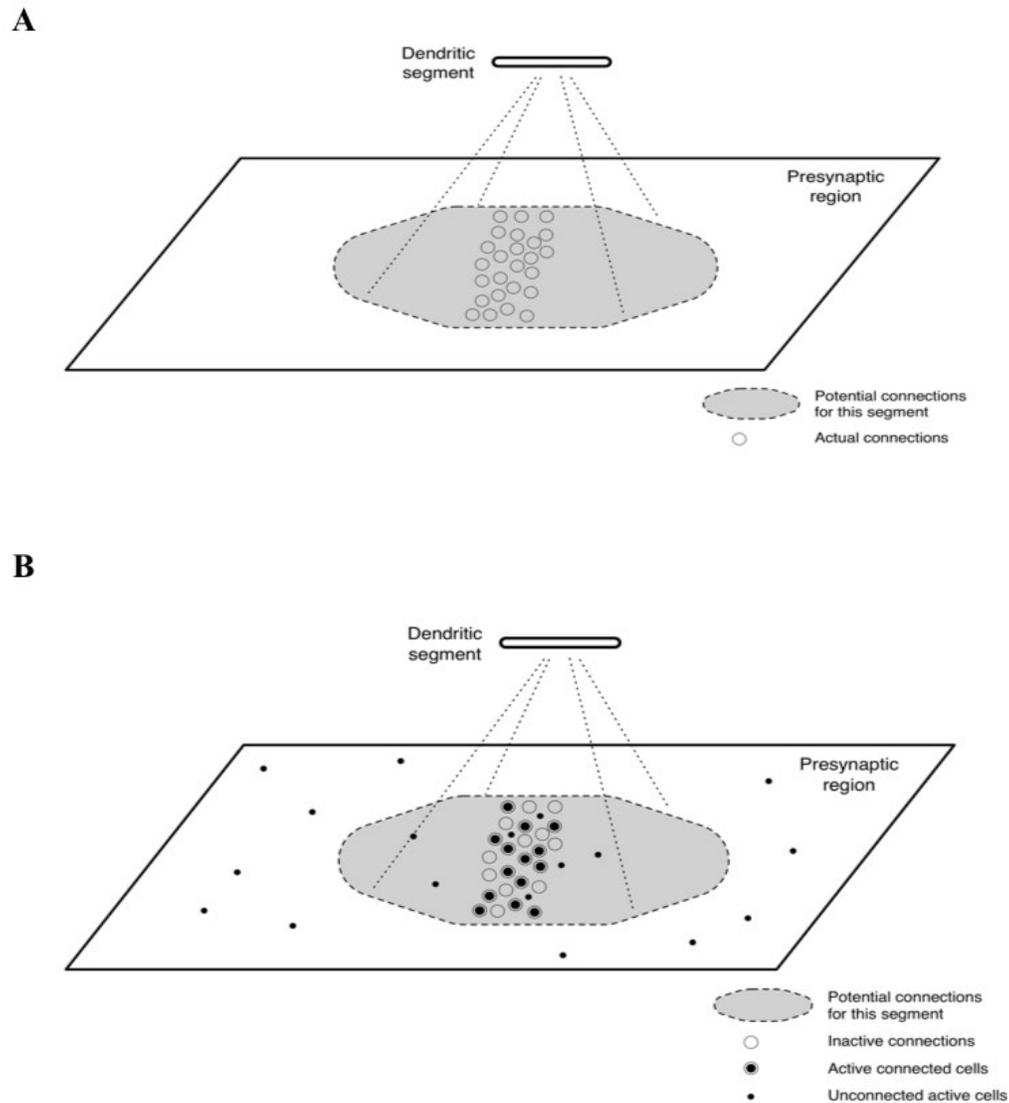

**Figure 2. A.** This figure shows a single example dendritic segment and its connections to some presynaptic area $A_t$. The grey area represents the set of potential connections to this segment. The clear circles denote 25 presynaptic cells corresponding to 25 synapses on the segment. In this example the synapses detect a roughly vertical bar of activity near the center of $A_t$. The bar is purely for illustration purposes - in general the segment could contain an arbitrary random subsample of $A_t$. **B.** This figure illustrates the behavior of the segment given a specific pattern of active presynaptic neurons. The small black circles represent the currently active cells in $A_t$, and filled circles denote synaptic connections that overlap with active cells. In this example there is a fair amount of noise. Many of the cells within the segment's preferred vertical pattern did not become active and there are several additional active cells. However 14 of the active cells are connected; a threshold of 14 or lower will cause the segment to initiate a dendritic spike. When the dimensionality is high, the chance that a random pattern of activity will overlap sufficiently with the segment is extremely low, hence it is very likely that the activity represents an actual vertical pattern, and not an accident.

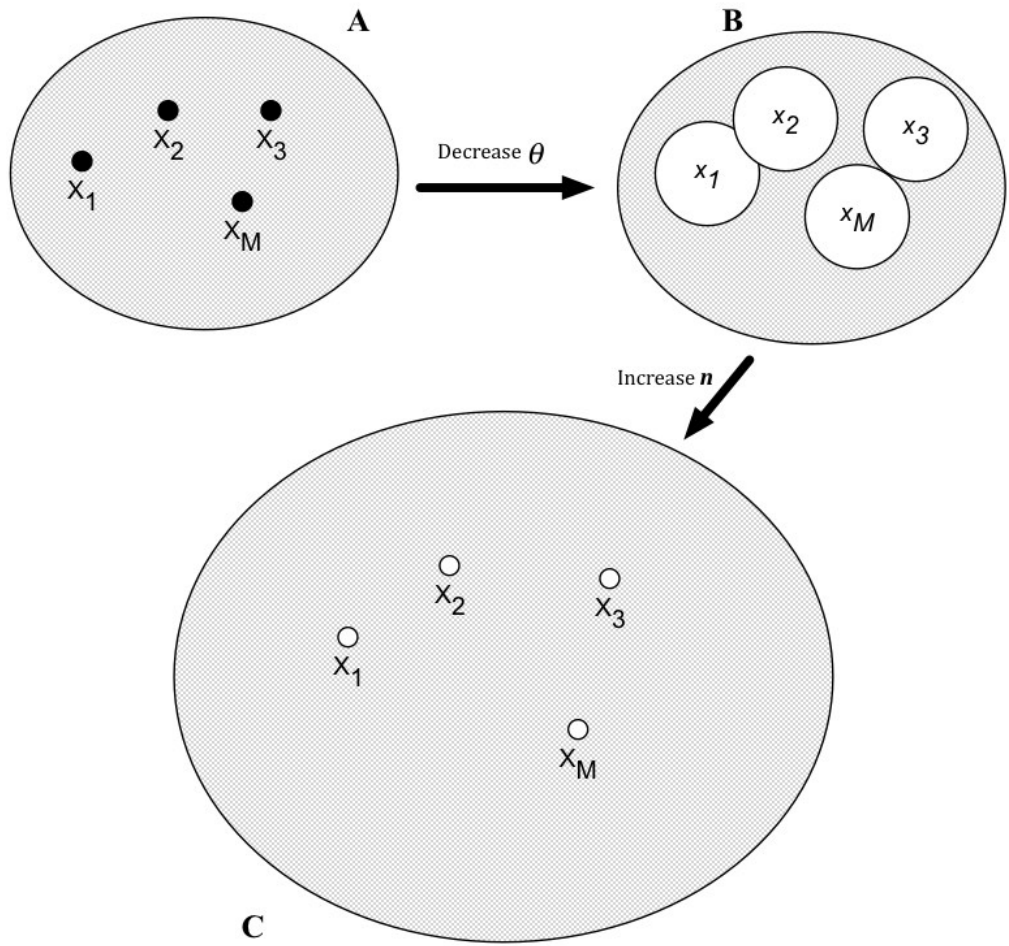

**Figure 3.** This figure illustrates the conceptual effect of decreasing $\theta$, the dendritic threshold and increasing $n$, the overall population of cells. The large grey ovals denote the universe of possible patterns. The smaller circles within indicate $M$ different dendritic segments, each representing one pattern within this universe. If $\theta$ is equal to the number of synapses $s$, as shown in oval A, very few patterns will match any of these segments (this is denoted by the small black circles). As you decrease $\theta$ the set of potential matches to each segment increases, as indicated by the large white circles in oval B. With smaller $\theta$ the segments will therefore be more tolerant to changes but there is a larger probability of false matches to random patterns. The ratio of white to grey becomes larger as you decrease $\theta$. Oval C shows that when you increase $n$, the universe of possible patterns increases and the relative size of the white circles shrinks rapidly. Thus there is a tradeoff between these parameters. You get more robustness as you decrease $\theta$ at the cost of additional false positives. This is mitigated if you increase the overall cell population $n$.

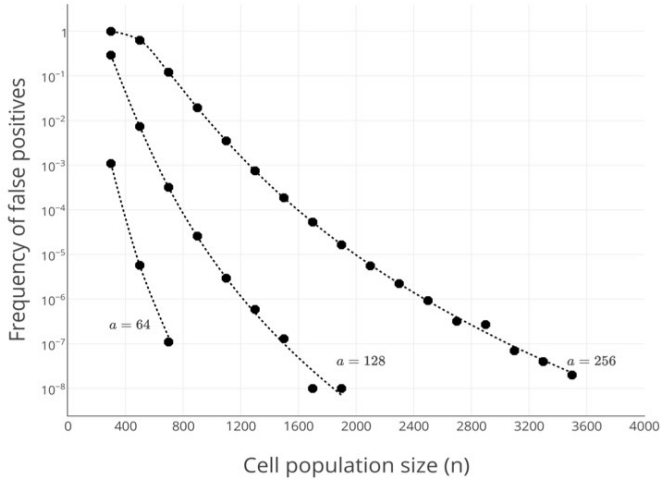

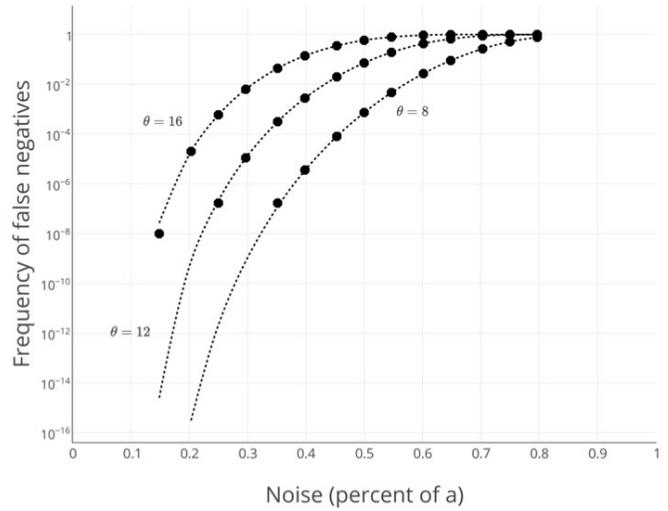

**A**                                **B**

**Figure 4. A.** This figure shows false positive rates as a function of presynaptic activity level, $a$, and the overall number of presynaptic cells, $n$. The dotted lines show theoretically predicted values (from Eq. (4)) whereas the black circles show the values observed through simulation. Each circle denotes match percentage based on 100 million comparisons between a randomly sampled model dendrite and random presynaptic activity. False positive rates drop rapidly as the underlying dimensionality increases. In this chart the number of synapses, $s$, was fixed at 24 and the dendritic threshold, $\theta$, was fixed at 12. **B.** This figure shows false negatives as a function of noise and dendritic threshold $\theta$. The dotted lines show theoretically predicted values (from Eq. (5)) whereas the black circles show the values observed through simulation. Each circle denotes the frequency of failed matches between a segment that stores a subsample $s$ from a presynaptic vector $\boldsymbol{A}$, and a noisy version of that vector, $\boldsymbol{A}^*$. A noise level of 0.1 indicates that $\boldsymbol{A}^*$ was created by swapping 10% of the active bits in $\boldsymbol{A}$ with other bits. 10 million random trials were used to calculate each point. In this chart the presynaptic activity level, $a$, was fixed at 128, $n$ at 6,000 and $s$ at 30. We do not display error bars in either figure, as they are negligible.

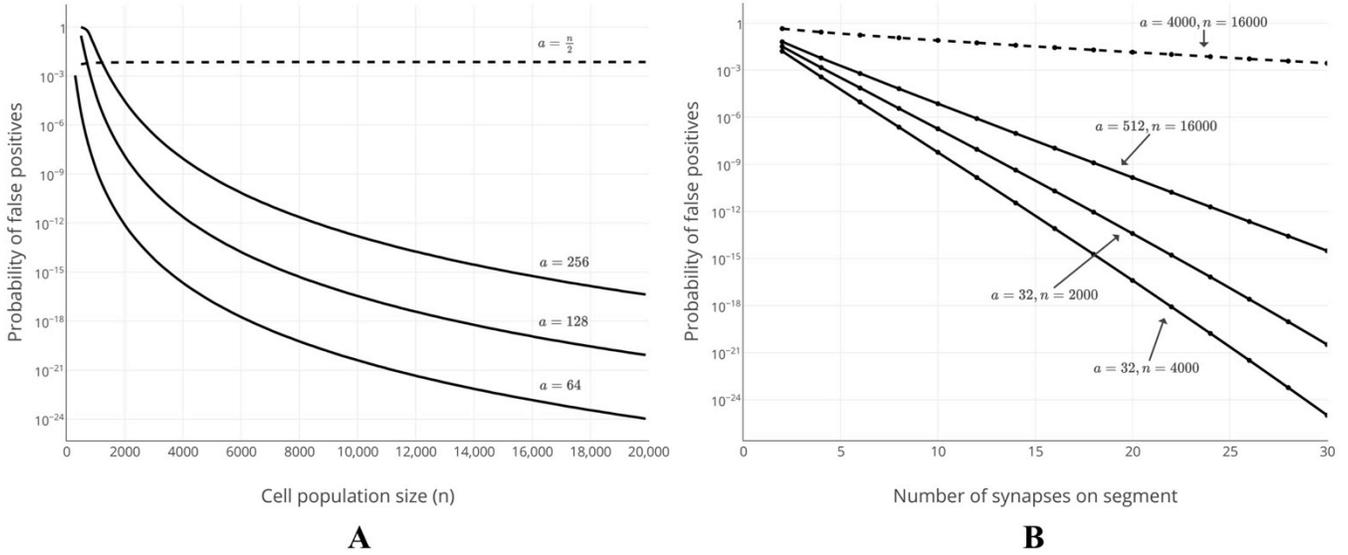

**A**                                   **B**

**Figure 5. A**. This graph illustrates the behavior of Eq. (4) and the effects of cell population and sparsity. The three solid curves show the rapid drop in error rates as the number of cells *n* increases. Each curve shows a different sparsity level. For example, if 128 out of 4000 cells are active (3.2% sparsity) the error rate is a little higher than $10^{-12}$. The dashed line corresponds to an activity level of 50%. The fact that the error corresponding to this condition does not drop demonstrates that both sparsity and high dimensionality are required to achieve low error rates. In all of these simulations, the number of synapses *s*=24 and the dendritic threshold $\theta$=12, corresponding to 50% noise tolerance. **B**. This graph illustrates the behavior of Eq. (4) and the effect of *s*, the number of synapses on a segment. The segment threshold is set to $\theta = \frac{s}{2}$. As the curves show, there is an exponential improvement in error rates as the number of synapses increases. The three solid curves show error rates for a variety of sparsity levels ($\frac{a}{n}$ is between 0.8% to 3.2%) and a range of dimensionalities (*n* between 2000 and 16000). The curves show that if the overall activity level is sparse and the cell population is sufficiently high, values of *s* between 15 and 25 can lead to low error rates. This helps explain why even a tiny number of synapses on a dendritic segment subsampling from a much larger pattern of activity is sufficient for robust recognition performance. The dashed curve is an example illustrating that a relatively dense representation (25% shown) does not lead to low error even if the underlying dimensionality is high.

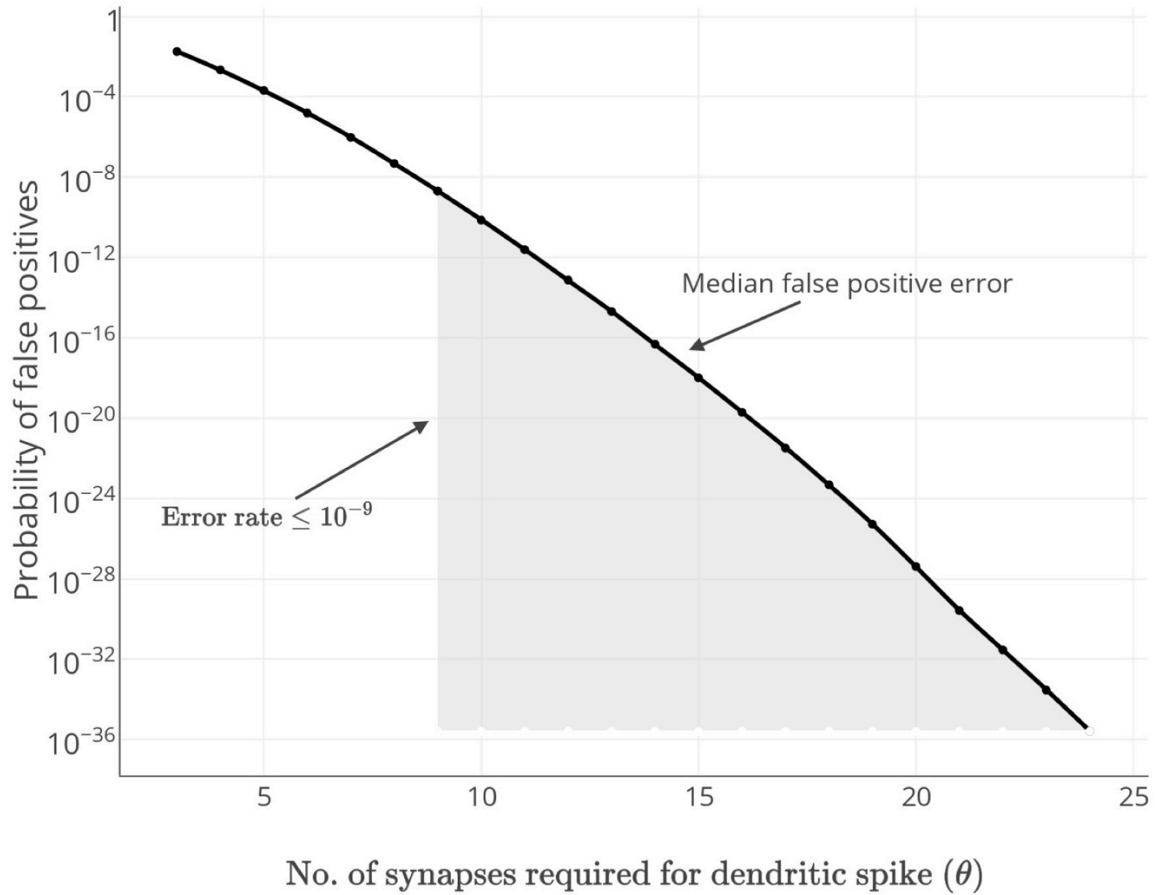

**Figure 6.** The equations can be used to characterize the effect of different synaptic thresholds and predict an ideal range. This chart shows the median probability of error as a function of the synaptic threshold for a dendritic spike $\theta$. Each point on the graph holds $\theta$ fixed and represents the median probability of error computed over a large range of all other parameters: $n$ (the number of potential synapses), presynaptic activity $a$, and $s$, the number of synapses on each segment. We exhaustively varied $n$ from 10,000 to 200,000, presynaptic activity from 0.5% to 3% of $n$, and $s$ from 20 to 50. The shaded area of the chart is the region corresponding to low error (for illustration purposes we show the range of thresholds that lead to an error of 1 in a billion or lower). The chart demonstrates that a spike threshold of 9 and higher leads to low error rates under a very wide range of assumptions. A spike threshold beyond 15 or 20 leads to diminishing returns. Beyond that the error rates are so low that the extra metabolic cost of additional synapses due to higher thresholds is not justified.

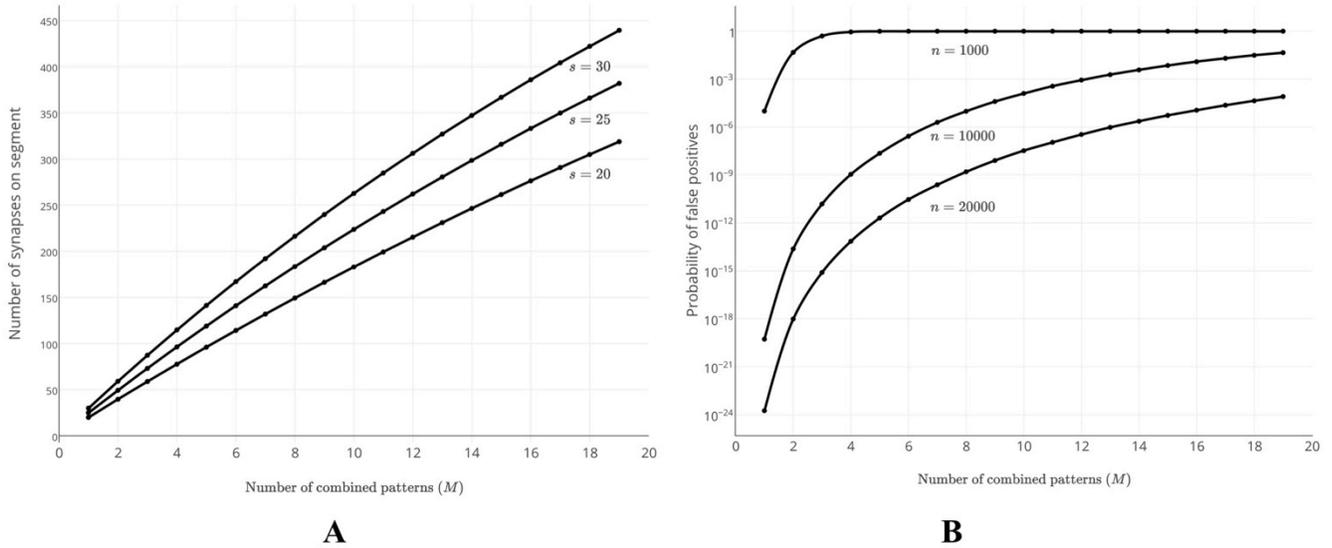

**A**    **B**

**Figure 7. A**. This figure plots the expected number of synapses on a segment that represents M different patterns. Each curve shows a different value of *s*, the number of synapses used represent a single pattern. The point of this plot is to show the curve grows slower than linearly because as more patterns are included in the union the chance that two patterns contain overlapping presynaptic cells grows. **B**. This figure shows the expected error rates for segments that contain a union of multiple patterns. The error increases monotonically with the number of patterns "stored" on each segment. The size of the presynaptic population is a critical factor; a larger presynaptic population implies a much lower probability for mix and match errors. A small presynaptic population of 1000 leads to relatively high error rates, but a population of 20,000 can lead to extremely low error rates even with 10 patterns stored. In this graph the activity of the population, *a*, is 200. *s*, the number of synapses representing each pattern, is 25. The dendritic threshold $\theta$ is 15.

23